\documentclass[
    ,final            % use final for the camera ready runs
  ]
  {aipproc}

\layoutstyle{6x9}
\usepackage{graphicx} % Include figure files

\begin{document}

\title{Sides of ``The'' Unitarity Triangle: \\ Results from Belle and Babar
\footnote{University of Cincinnati preprint \# UCHEP-05-05}
\footnote{Talk presented at XXV Physics in Collision, 6-9 July, 2005, Czech Republic}
}


\author{K. Kinoshita}{
  address={Department of Physics\\University of Cincinnati\\Cincinnati, OH 45221 USA}
}

\begin{abstract}
The principal goal of the $B$-factories is to test the CKM paradigm through measurements that overconstrain the shape of the so-called Unitarity Triangle.  
The sides of the triangle are evaluated through absolute values of several elements of the CKM matrix, so achieving high precision on these is an important aspect of the $B$-factory program.  
 Recent results from Belle and Babar are presented.  
\end{abstract}

\maketitle

%%%%%%%%%%%%%%%%%%%%%%%%%%%%%%%%%%%%%%%%%%%%
%% MAINMATTER
%%%%%%%%%%%%%%%%%%%%%%%%%%%%%%%%%%%%%%%%%%%%

\section{Introduction}

The weak charged-current couplings of the quarks, arranged in the $3\times 3$ matrix known as the CKM matrix,
describe in the Standard Model the transformation between the mass and weak eigenstates of the quarks.
As a transformation between two complete sets of eigenstates, the matrix must be complex as well as preserving the orthogonality and metric, {\it i.e.}, it must be unitary.   
Formally, the elements $\{V_{ij}\}$ of a unitary matrix must satisfy
$\sum_{j}V_{ji}^*V_{jk}=\delta_{ik}$,
and for a $3\times 3$ matrix the constraints imposed by these conditions reduce the freedom of CKM to three real and one irreducibly complex parameters, often represented explicitly as
\begin{eqnarray*}
{\cal M}=
\left(\begin{array}{ccc}
V_{ud}&V_{us}&V_{ub}\\
V_{cd}&V_{cs}&V_{cb}\\
V_{td}&V_{ts}&V_{tb}
\end{array}\right)
=
\left(\begin{array}{ccc}
1-\lambda^2/2& \lambda  & \lambda^3A(\rho -i\eta)\\
-\lambda & 1-\lambda^2/2 & \lambda^2 A\\
\lambda^3A(1-\rho -i\eta) & -\lambda^2 A & 1
\end{array}\right)
\end{eqnarray*}
The unitarity condition, applied to $\{i=1,k=3\}$, results in
\begin{eqnarray}
0&=&{V_{ub}^*V_{ud}\over V_{cb}^*V_{cd}}+1+ {V_{tb}^*V_{td}\over V_{cb}^*V_{cd}}\label{eq:UT}\\
&\approx& -(\rho+i\eta)+1-(1-\rho-i\eta).\nonumber
\end{eqnarray}
This sum of three terms may be represented as a closed triangle in the complex plane with corners at $(0,0)$, $(1,0)$ and $(\rho,\eta)$.
In the context of the $B$-factories, this has come to be known as ``{\it The} Unitarity Triangle'' and embodies the least precisely known aspects of the CKM matrix.
The objective of the $B$-factory experiments is to test the validity of Equation~(\ref{eq:UT}), {\it i.e.} the closure of this triangle.
While much attention has been focused on the angles $\phi_1(\beta)$, $\phi_2(\alpha)$, and $\phi_3(\gamma)$, which are measured through $CP$ asymmetries, any three among the three angles and sides fully constrain the triangle, so the precise measurement of the sides is an important aspect of the $B$-factory program.
Of the CKM elements involved, $|V_{ub}|$, $|V_{cb}|$, and $|V_{td}|$ address this issue directly and are accessible to the $B$-factories.

The absolute value of a coupling, $V_{xx}$, is probed via decays that are represented dominantly by a single decay diagram, where all other coupling and processes are well known such that the decay rate may be expressed as the product of $|V_{xx}|^2$ and a known quantity.
For example, each of the decays reported here proceeds by one of the two mechanisms, ``tree'' or ``penguin,'' shown in the figure.
\begin{figure}[ht]
  \includegraphics[height=3.cm]{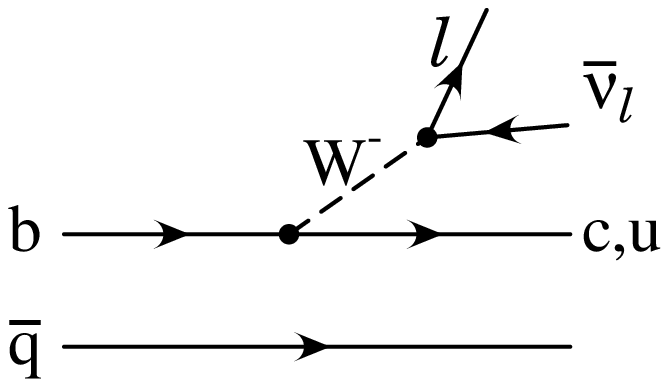}\hfill
  \includegraphics[height=3.cm]{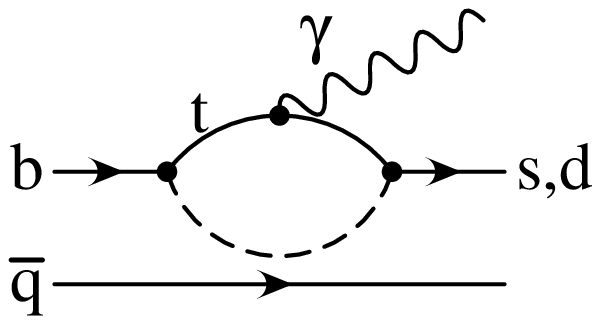}
  \caption{B decay processes used to probe $|V_{cb}|$, $|V_{ub}|$, and $|V_{td}|$: tree (left) and penguin (right).}
  \label{fig:decays}
\end{figure}
A relevant feature of both is that a minimumal number of valence quarks is involved;
large uncertainties associated with hadronic states are thus minimized (although it is not possible to avoid them entirely).

Progress in both experiment and theory is resulting in continuous improvements to our knowledge of $|V_{ij}|$.
On the experimental front, the $B$ factories continue to accumulate data at a high rate, with 466~fb$^{-1}$ accumulated at Belle and 262~fb$^{-1}$ at Babar, providing access to clean but rare modes and to corners of phase space where theoretical uncertainties are minimal.
From the theoretical side, new approaches and refinements as well as the arrival of unquenched lattice QCD results have enabled the kind of precision that would have been considered optimistic a decade ago.

\section{Results on and Status of $|V_{\lowercase{td}}|$}
While $B_s$ mixing is better known in the context of measuring $|V_{td}|$, this coupling is also approachable in principle by the radiative process $b\to d\gamma$, shown in Figure~\ref{fig:decays}(right).  
As with $B_s$ mixing,  taking a ratio of rates for the targeted mode and for a more easily measured mode that differs by a single coupling, $b\to d\gamma$ and $b\to s\gamma$ in this case, results in the cancellation of much theoretical uncertainty. 
%\begin{eqnarray*}
%$
%{\Gamma(b\to d\gamma)\over \Gamma(b\to s\gamma)}\propto
%\left|{V_{td}\over V_{ts}}\right|^2.
%$
%\end{eqnarray*}

\begin{figure}[t]
\includegraphics[width=0.8\textwidth]{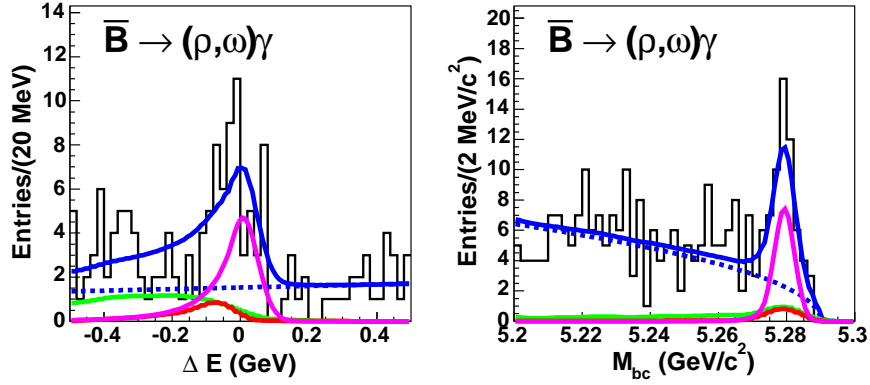}
  \caption{Candidates for $\bar B^0\to\rho^0\gamma$, $B^-\to\rho^-\gamma$, and $\bar B^0\to\omega\gamma$ (combined) in 386 million $B\bar B$ events from Belle. The data are shown with all modes combined, for display purposes only.}
  \label{fig:b2drad_data}
\end{figure}

Both Belle and Babar have searched for the exclusive modes $B\to\rho\gamma$ and $B\to\omega\gamma$.  
In 386M $B\bar B$ events, Belle has found evidence for $b\to d\gamma$ in the modes $\bar B^0\to\rho^0\gamma$, $B^-\to\rho^-\gamma$, and $\bar B^0\to\omega\gamma $  \cite{btod}.
The modes are identified through full reconstruction of 
decays in  $e^+e^-\to\Upsilon$(4S)$\to B\bar B$ events, where
backgrounds are tremendously suppressed due to the well-defined energy and low momentum of the $B$'s in the $e^+e^-$  center-of-mass ({\it cms}).  
These properties are characterized by two variables, $\Delta E\equiv E^*_{cand}-E^*_{beam}$ and $M_{\rm bc}\equiv\sqrt{E^{*2}_{beam}-p^{*2}_{cand}}$, where the superscript ($^*$) denotes a quantity in the {\it cms}. 
Figure {\ref{fig:b2drad_data}} displays candidate distributions in $\Delta E$ and $M_{\rm bc}$ for all modes combined.
Under an assumption of isospin invariance, 
${\cal B}(B^-\to\rho^-\gamma)=2{\tau_{B^+}\over\tau_{B^0}}{\cal B}(\bar B^0\to\rho^0\gamma)=2{\tau_{B^+}\over\tau_{B^0}}{\cal B}(B^0\to\omega\gamma)$, the three modes are combined into a single measurement of ${\cal B}(B^-\to\rho^-\gamma)$ (designated as ${\cal B}(B\to (\rho/\omega)\gamma)$) at 5.5$\sigma$ significance:
\begin{eqnarray*}
{\cal B}(B\to (\rho/\omega)\gamma)&=&(1.34{^{+0.34}_{-0.31}}{^{+0.14}_{-0.10}})\times 10^{-6}
\end{eqnarray*}
The ratio with   the corresponding $b\to s\gamma$ value,
\begin{eqnarray*}
{{\cal B}(B^-\to (\rho/\omega)\gamma)\over {\cal B}(B^-\to K^{*-}\gamma)}&=&0.032{\pm 0.008}{^{+0.003}_{-0.002}},\\
%\end{eqnarray*}
{\rm yields}\ \ \ \ \ \ \ \ \ \ \ \ \ \ \ \ \ \ \ \ 
%\begin{eqnarray*}
\left|{V_{td}\over V_{ts}}\right|&=&0.200{^{+0.026}_{-0.025}}{^{+0.038}_{-0.029}}.
\end{eqnarray*}
Babar's search for the same modes among 211M~$B\bar B$ events\cite{ref5} yielded possible signals at the 2.1$\sigma$ level and the following 90\% confidence upper limits:
\begin{eqnarray*}
{\cal B}(B^-\to (\rho/\omega)\gamma)<1.2\times 10^{-6},\ 
{{\cal B}(B^-\to (\rho/\omega)\gamma)\over {\cal B}(B^-\to K^{*-}\gamma)}<0.029,\ 
\left|{V_{td}\over V_{ts}}\right|<0.19.
\end{eqnarray*}

\section{Results on and Status of $|V_{\lowercase{cb}}|$}

As no new results have been released since the Winter conferences of March 2005, I will simply present a very brief status summary.
Two principal methods, both using semileptonic $b\to c\ell\bar\nu$ decays, are used to evaluate $|V_{cb}|$.
For a number of years after the development of Heavy Quark Effective Theories (HQET)\cite{hqet}, the highest precision was obtained with the modes $B\to D^{*}\ell\bar\nu$.  
Exclusive modes are particularly clean theoretically if the final state charm particle is at rest with respect to the $B$ parent; the form factor is unity if $b$- and $c$-quarks are infinitely massive, with minor corrections for finite mass heavy quarks.  
This method has experimental challenges, both in terms of statistics and control of systematics.
The current best values from exclusive decays, which can be seen at the Heavy Flavors Averaging Group (HFAG) website,
\{http://www.slac.stanford.edu/xorg/hfag/semi/winter05/winter05.shtml\},
are 
\begin{itemize}
\item
from $B\to D^*\ell\bar\nu$: \\
${\cal F}(1)|V_{cb}|=(37.6\pm 0.9)\times 10^{-3}$; \\
using ${\cal F}(1)=0.91\pm 0.04,\ |V_{cb}|=(41.3\pm 1.0\pm 1.8)\times 10^{-3}$.
\item
from $B\to D\ell\bar\nu$: \\
${\cal G}(1)|V_{cb}|=(42.2\pm 3.7)\times 10^{-3}$; \\
using ${\cal G}(1)=1.04\pm 0.06,\ |V_{cb}|=(40.6\pm 3.6\pm 1.3)\times 10^{-3}$.
\end{itemize}
where ${\cal F}(1)$ and ${\cal G}(1)$ are the respective form factors at zero recoil.

More recently, $|V_{cb}|$ has been evaluated through moments of leptonic momentum and hadronic mass distributions from inclusive $b\to c\ell\bar\nu$ decays, and this method currently achieves the highest precision.
Both Babar\cite{babar_moments} and Belle\cite{belle_0408139} have reported measurements of moments, requiring one fully reconstructed hadronic $B$ decay in each event and measuring the products of a semileptonic decay among the remaining detected particles.
From 89M $B\bar B$ events, Babar reports $|V_{cb}|=(41.4\pm 0.4\pm 0.4\pm 0.6)\times 10^{-3}$.
An update and report of $|V_{cb}|$ from Belle is in progress.
A global fit analysis  of results from Babar, Belle, CDF, CLEO, and DELPHI under a common theoretical framework was performed by Bauer {\it et al.} \cite{bauer}, to yield $|V_{cb}|=(41.4\pm 0.6\pm 0.1)\times 10^{-3}$, where the first error is from the fit and the second is due to the uncertainty on the $B$ lifetime.

An excellent summary of details on the measurement of $|V_{cb}|$ is presented in Ref.~\cite{artuso}.

%\newpage
\section{Results on and Status of $|V_{\lowercase{ub}}|$}
The history of $|V_{ub}|$ has been one of  theoretical more than experimental limitations. 
Nonetheless, much progress has been made in recent years, due in large part to the maturing of the $B$ factories and their huge available data samples.
The new capabilities presented by these data have not only inspired theorists to refine their models, but have also stimulated dialogues between experimenters and theorists and improved the targeting of key issues.

The earliest measurements examined the  inclusive lepton spectrum for decays of the type $b\rightarrow u\ell^-\bar\nu$ at energies above the kinematic limit for $b\to c\ell\bar\nu$, the so-called ``endpoint'' analysis.
This method yields relatively high statistics, but the precision on  $|V_{ub}|$ was limited theoretically by our imprecise knowledge of hadronic wavefunctions, of the $B$ and of the low-mass final states which contribute dominantly to the endpoint region.

While most analyses still target decays of the type $b\rightarrow u\ell^-\bar\nu$, there has been a proliferation of approaches.
The main issues remain the rarity of $b\to u\ell\bar\nu$ in the face of dominant $b\to c\ell\bar\nu$ backgrounds and the necessarily incomplete reconstruction of decays that include a neutrino.
With substantial samples of $b\to u\ell\bar\nu$ decays, experimenters have been able to dissect kinematic distributions, resulting in reduced model dependence, improved signal significance, and improved robustness for $|V_{ub}|$.
We are on the verge of being able to make meaningful comparisons between different theoretical models.
In addition, exclusive modes are now contributing in a significant way, thanks to new results from lattice and light-cone sum rule calculations.

\subsection{Inclusive $b\to u\ell\bar\nu$ decays}
The various methods of identifying inclusive $B\to X_u\ell\bar\nu$ decays contribute with varying balances of statistical power and robustness against systematic experimental and theoretical uncertainties.
While the lepton is used in all methods, one may also obtain an indirect measurement of the neutrino via missing energy and momentum as well as a measurement on the hadronic part (``$X_u$'') of the final state using one of several methods to separate the associated hadrons from the products of the other $B$ in the event.
The aim in obtaining information on the neutrino and/or $X_u$ is twofold: 
\begin{itemize}
\item
to obtain distributions in the more theoretically relevant quantities, such as $q^2$ and $M_X$ as well as $P_+\equiv E_X-P_X$ so that model-dependence is greatly reduced,
\item
to optimize the reduction of background from $b\to c\ell\bar\nu$ and allow exploration of a larger portion of the phase space, thereby reducing theoretical errors.  
\end{itemize}
The cleanest measurements are made by tagging with a fully reconstructed hadronic $B$ decay (at a cost of >10$^2$ in statistics) which, in addition to removing combinatorial background, provides a measurement of the $B$'s 3-momentum.

The extraction of $|V_{ub}|$ from a measured partial rate requires theoretical input, and this area has seen much recent activity.  
The result of Lange, Neubert, and Paz \cite{lnp} (LNP) is applied in the most current reports.
This work has two salient features: it provides for the direct calculation of $|V_{ub}|$ from a measured partial rate without (model-dependent) extrapolations, and it operates within the framework of the ``shape function'' scheme  \cite{neubert_sf}.
This scheme has been stimulated by the emergence of rare $B$ decays as a probe of hadronic structure;
in particular, 
the energy of the photon from a $B\to X_s\gamma$ decay (in the $B$ center-of-mass) fixes the mass $M_X$ of the recoiling hadronic jet so that the photon spectrum is a direct probe of the internal structure of the $B$ meson, characterized by a ``shape function.''
The corresponding structure of $B\to X_u\ell\bar\nu$ decays may be represented in terms of the same shape function, so that in principle the determination of the measured spectrum from $B\to X_s\gamma$ can be used to specify the structure of $B\to X_u\ell\bar\nu$ and relate its rate to $|V_{ub}|$.

\begin{figure}
  \includegraphics[height=5.cm]{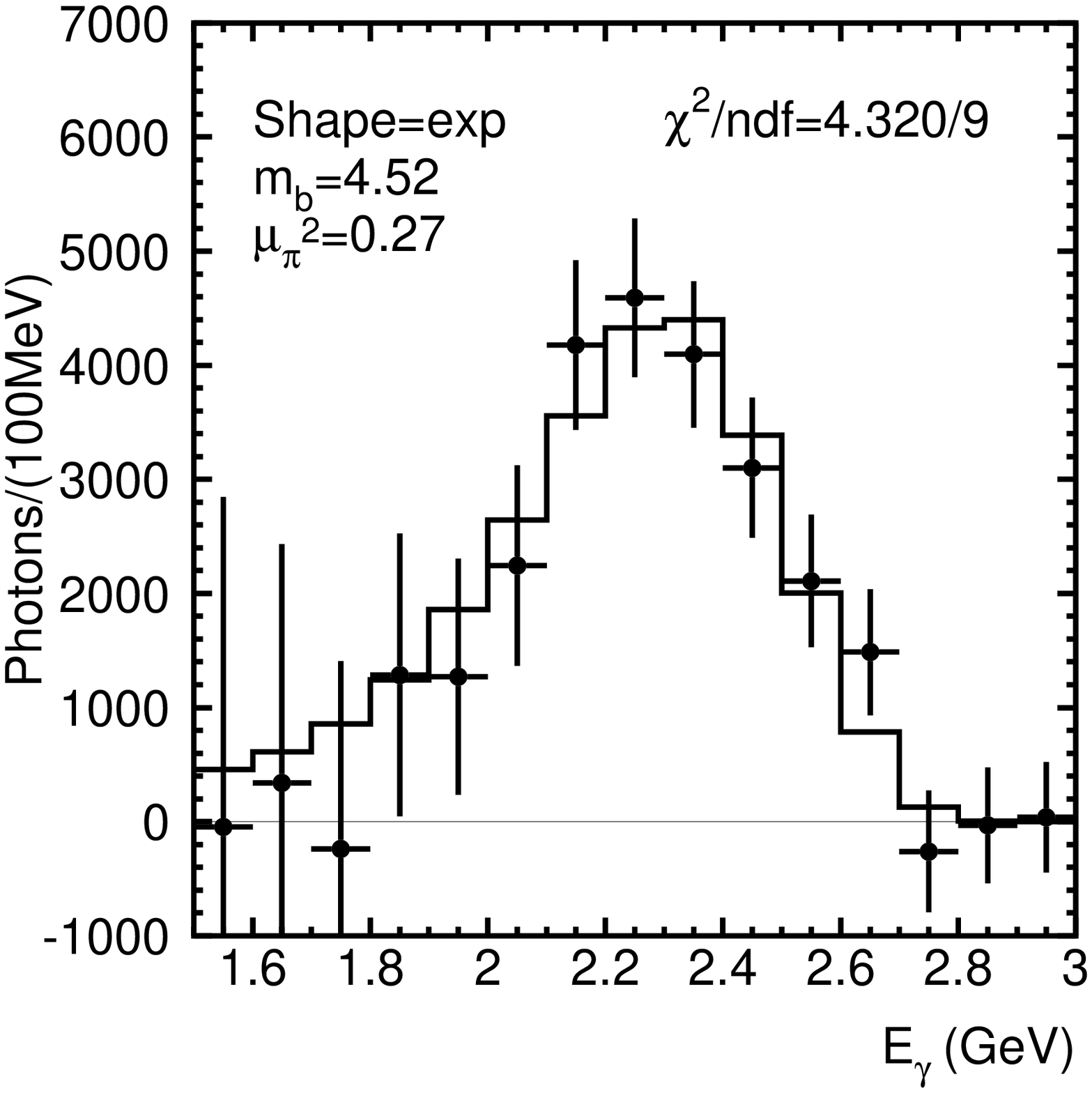}
  \includegraphics[height=5.cm]{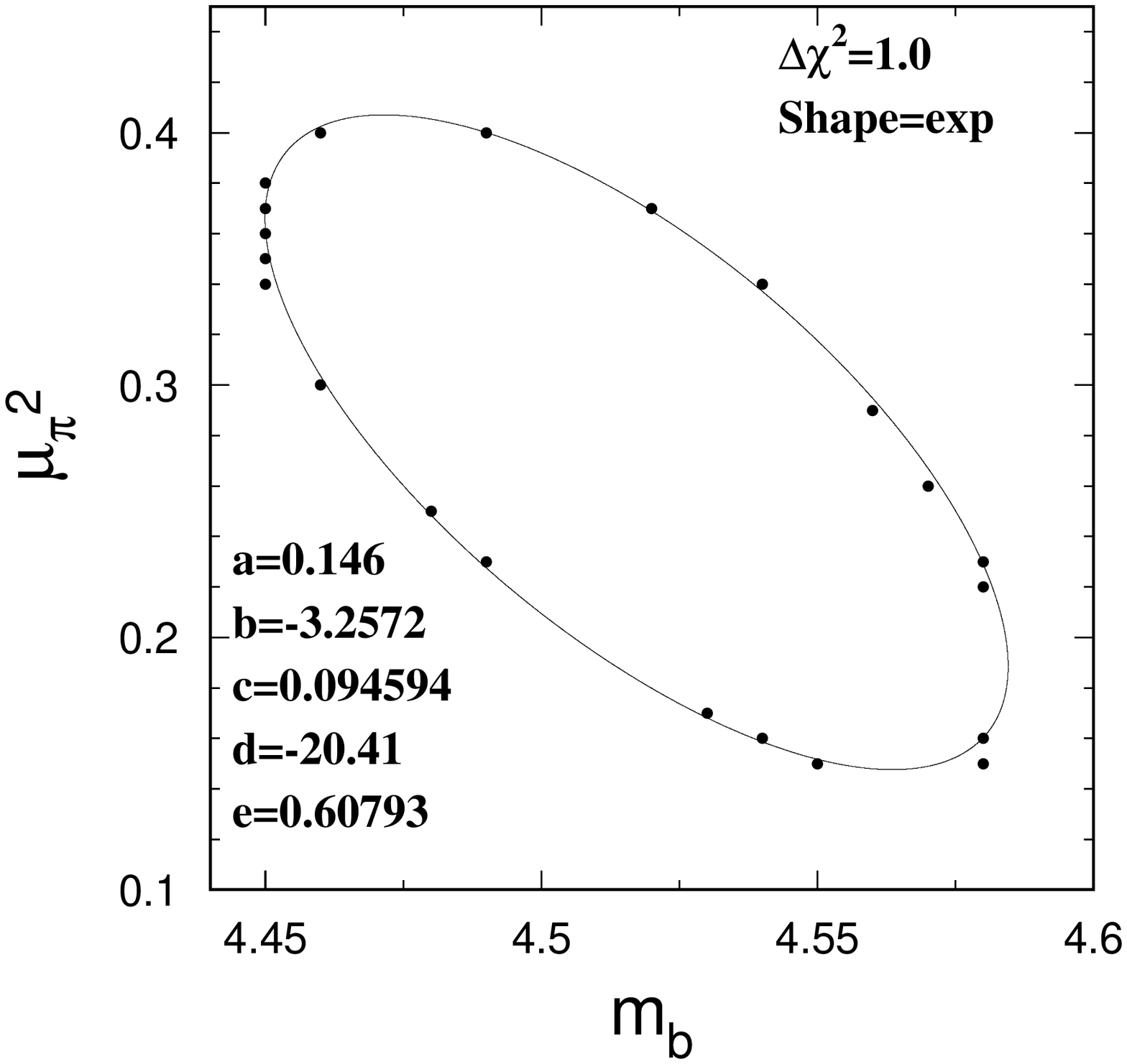}
\includegraphics[height=5.cm]{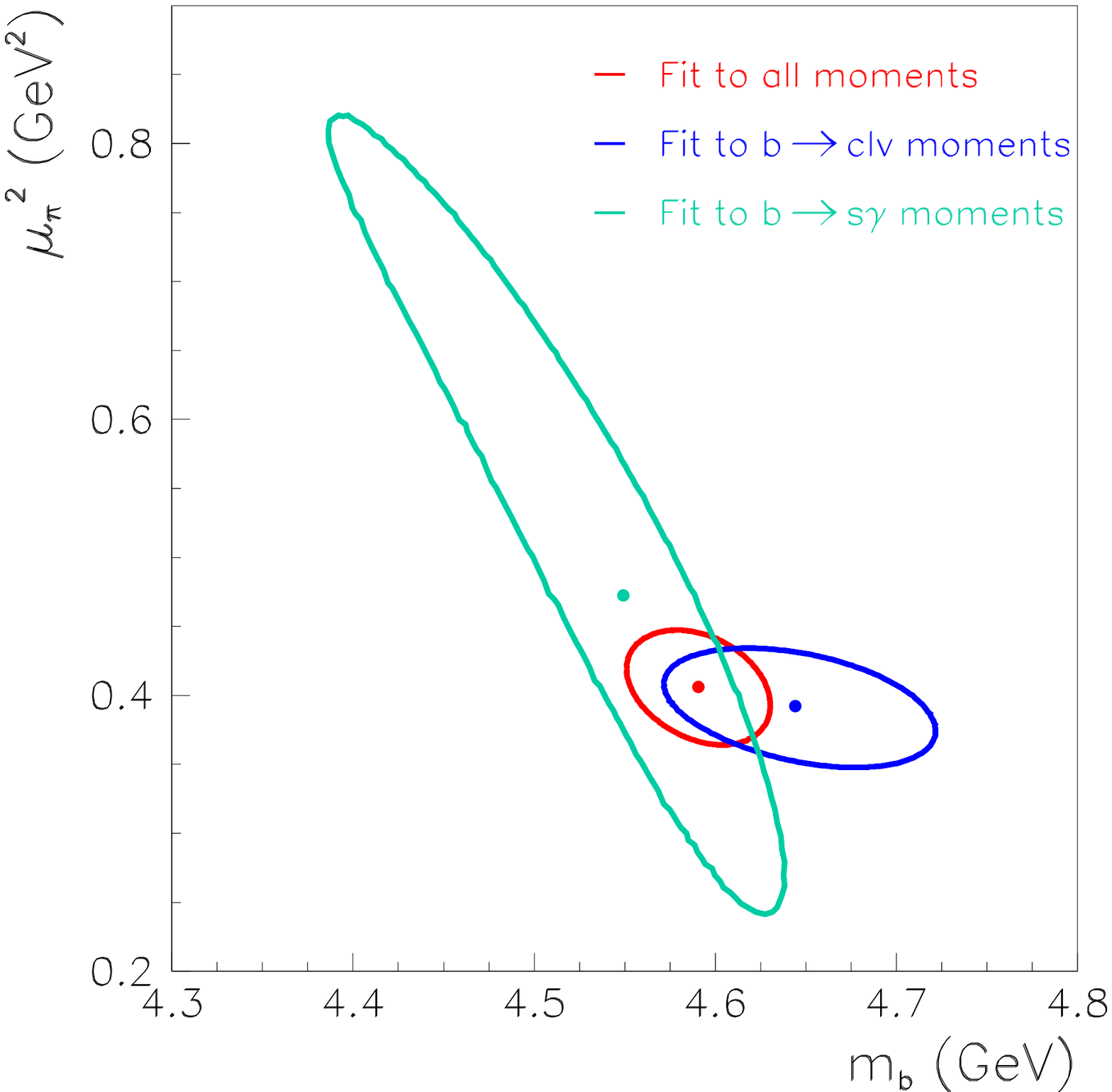}
\caption{(Left) Inclusive photon spectrum from decays $B\rightarrow X_s\gamma$, after background subtractions (Belle\protect\cite{belle_0506057}). (Center) 1$\sigma$ error ellipse on shape function parameters $m_b$(SF) and $\mu_\pi^2$(SF) under the shape function scheme. (Right) from HFAG: 1$\sigma$ error ellipse on parameters $m_b$ and $\mu_\pi^2$ under the kinetic scheme.}
  \label{fig:shapefun}
\end{figure}

Reported results have used various forms for the shape function to be used as input to LNP.
In the most direct method, a parametrized form is fitted to the measured $B\to X_s\gamma$ spectrum to obtain two heavy quark parameters, the effective mass of the $b$ quark, $m_b$, and the mean square momentum of the $b$ quark, $\mu_\pi^2$.
The same form is then used to obtain $|V_{ub}|$.
The most recent fit to the Belle $B\to X_s\gamma$ photon spectrum (Figure \ref{fig:shapefun}(left)) yields  $m_b$(SF)$=4.52\pm 0.07~{\rm GeV}/c^2$ and $\mu_\pi^2$(SF)$=0.27\pm 0.13~{\rm GeV}^2/c^2$\cite{belle_0506057}, with the 1$\sigma$ error ellipse shown in Figure \ref{fig:shapefun}(center).

In a different approach, the ``kinetic'' scheme\cite{gambino}, the structure of the decays $B\to X_c\ell\bar\nu$, $B\to X_u\ell\bar\nu$, and $B\to X_s\gamma$ is based on a theoretical model, with common parameters $m_b$ and $\mu_\pi^2$
(where  $\mu_\pi^2$ is not the same as that for the shape function scheme).
The two parameters may be extracted from data on $B\to X_c\ell\bar\nu$  and/or $B\to X_s\gamma$.  
The resulting values are then translated to the corresponding parameters of the shape function scheme, establishing the form of the function.
While the determination of parameters in the kinetic scheme benefits from the high statistics of $B\to X_c\ell\bar\nu$, there has been some debate on the reliability of using charm modes in this application.
Results of fits in both shape function and kinetic schemes to $B\to X_s\gamma$ spectrum have been reported by Babar \cite{shape_babar}.

The endpoint analyses have recently been improved to allow the lower momentum cutoff to be reduced, to 1.9~GeV/c for Belle \cite{limosani} and 2.0~GeV/c for Babar \cite{endpt_babar}.
The determination of $|V_{ub}|$ from the endpoint has been further improved by the heavy quark parameter fits discussed above;
Belle has used a parametrization under the shape function scheme fitted to $B\to X_s\gamma$ results from Belle\cite{belle_0407052}, and Babar has used the kinetic scheme fitted to $B\to X_c\ell\bar\nu$ results from Babar\cite{shape_clnu}.
To extract $|V_{ub}|$, Belle uses {\cite{lnp}} while Babar has used \cite{kinetic_vub}.

\begin{figure}[b]
  \includegraphics[height=5.cm]{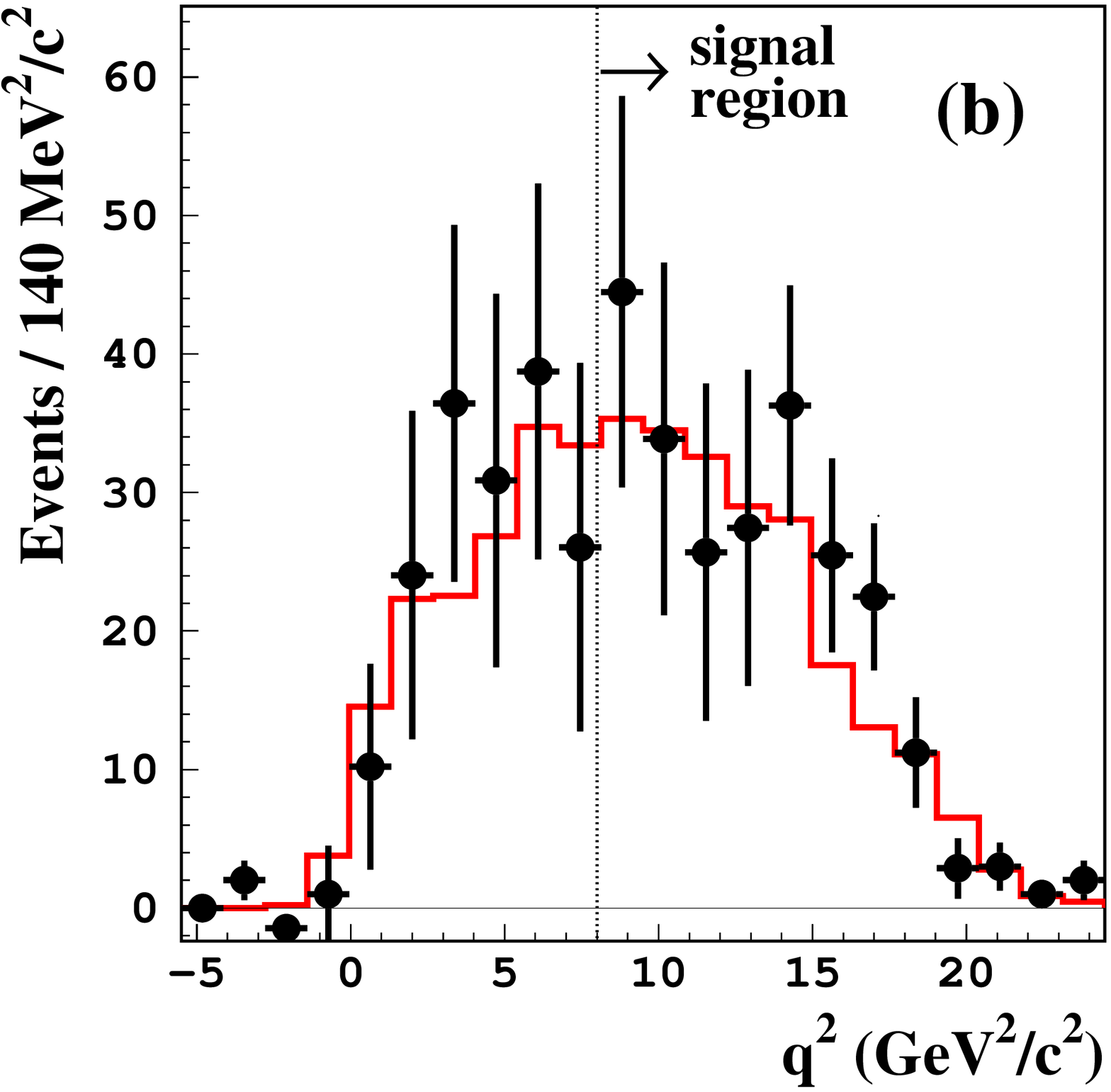}
  \includegraphics[height=5.cm]{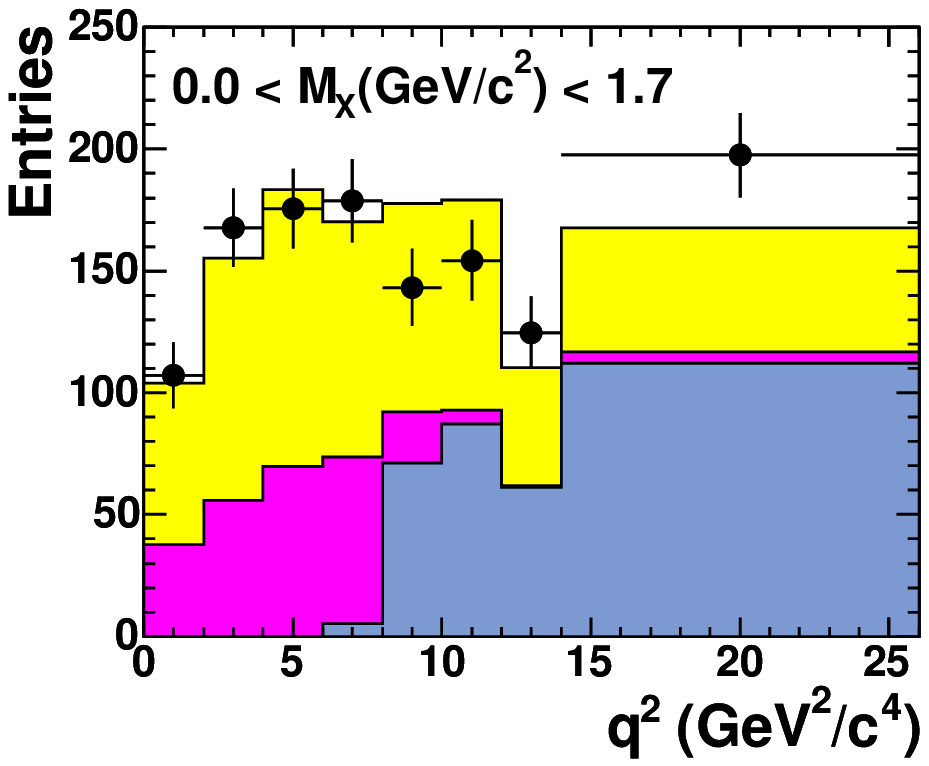}
\caption{Distribution in $q^2$ of $B\to X_u\ell\bar\nu$ candidates where the $b\to u$ fraction has been enhanced by requiring $M_X<1.7$~GeV/c$^2$, from Belle (left)\protect\cite{bizjak} and Babar (right) \protect\cite{b2u_FR_babar}.}
  \label{fig:partialrates}
\end{figure}

\begin{figure}
\includegraphics[height=5.cm]{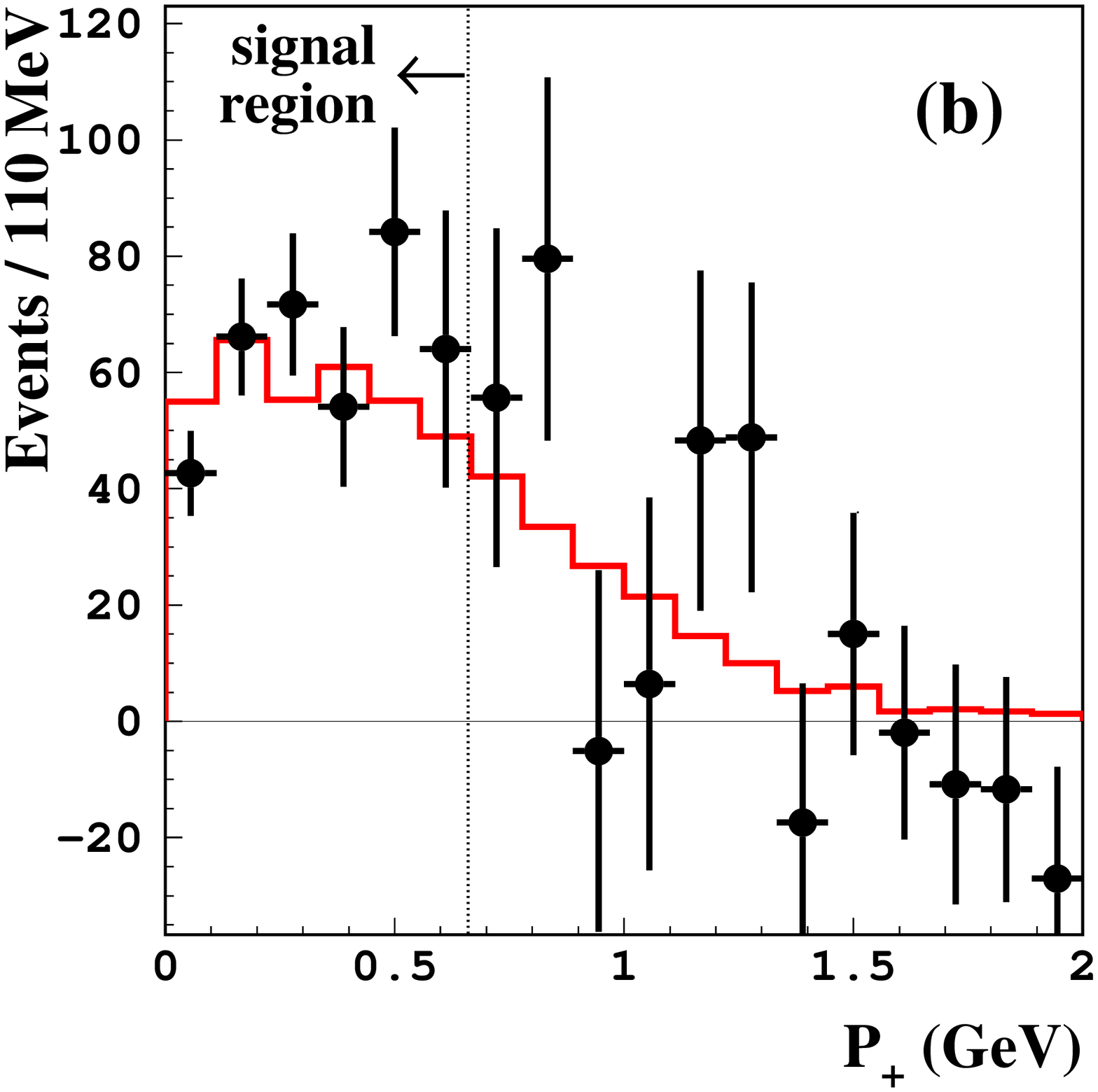}
\includegraphics[height=5.cm]{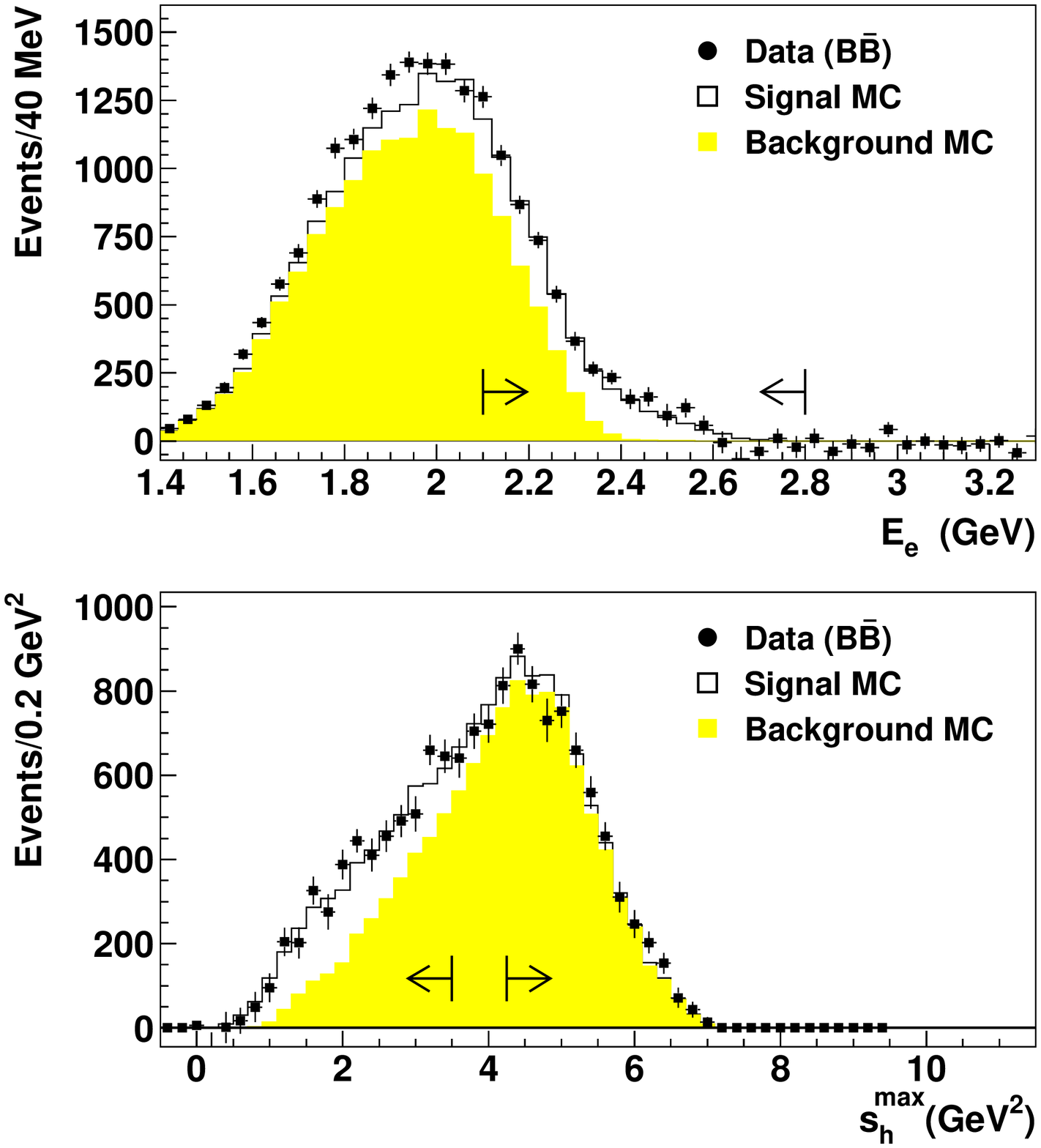}
\caption{(left) Distribution in $P_+$ of $B\to X_u\ell\bar\nu$ candidates after subtraction of backgrounds (points with errors) and prediction (histogram), from Belle \protect\cite{bizjak}. $P_+<0.66$ defines the signal region. 
(right) in Distributions of $B\to X_u\ell\bar\nu$ candidates in electron energy (top) and  \{maximum hadronic recoil mass\}$^2$ (bottom) in $\Upsilon$(4S) frame, from Babar \protect\cite{b2u_nurec_babar}.  }
  \label{fig:P+}
\end{figure}

Theoretical uncertainties may be reduced further through more kinematically detailed measurements, event-by-event. 
Both Babar and Belle report results achieved through full reconstruction tagging of one hadronic $B$ decay with examination of the lepton and hadronic system ``$X$'' in the residual event.  
Higher signal purity in these analyses allows acceptance of events with lower momentum leptons.
Close interactions between theory and experiment have resulted in theoretical calculations of rates in restricted phase space regions that balance theoretical robustness with experimental measurability;
cuts are targeted to minimize theory uncertainty while maximizing signal relative to the dominant $b\to c\ell\bar\nu$ backgrounds.
Figure~\ref{fig:partialrates} shows $q^2$ spectra from Belle (left) \cite{bizjak} and Babar (right)  \cite{b2u_FR_babar} of $B\to X_u\ell\bar\nu$ candidates where the $b\to u$ fraction has been enhanced by requiring low $M_X$.
Belle has also displayed candidate events in the variable $P_+\equiv E_X-P_X$, as recommended in {\cite{blnp}}, for improved separation from $b\to c\ell\bar\nu$ background while minimizing theory uncertainties (Figure~\ref{fig:P+}(left)).

Babar has also reported a measurement using electrons near the endpoint in combination with ``reconstructed'' neutrinos\cite{b2u_nurec_babar}.  
Figure~\ref{fig:P+}(right) shows the distributions of data and estimated background in the two variables used, electron energy and \{maximum hadronic recoil mass\}$^2$.  

\begin{table}[ht]
\begin{tabular}{lccc}
\hline\hline
{\bf Belle} & $\Delta{\cal B}\times 10^4$& $|V_{ub}|\times 10^3$\\ \hline
Endpoint (29M $B\bar B$)  \cite{limosani} & &\\ 
 $1.9<p_\ell<2.6$ GeV/c&   $8.47\pm 0.37\pm 1.53$\ &\ $5.08\pm 0.47\pm 0.49$\\ \hline
Full reconstruction tag ($p_\ell^*>1$~GeV/$c$) &\\
(275M $B\bar B$) \cite{bizjak} && \\
$M_X<1.7$~GeV/$c^2$, $q^2>8$~GeV$^2/c^2$ & $8.41\pm 1.14\pm 0.69$&$4.93\pm 0.33\pm 0.57$\\
$M_X<1.7$~GeV/$c^2$ & $12.4\pm 1.5\pm 0.8$&$4.35\pm 0.25\pm 0.46$\\
$P_+<0.66$~GeV/$c$ & $11.0\pm 1.5\pm 1.2$&$4.56\pm 0.30\pm 0.59$\\
\hline\hline 
{\bf Babar} &&\\ \hline
Endpoint (88M $B\bar B$)  \cite{endpt_babar} & &\\
$2.0<p_\ell^*<2.6$ GeV/c   & $4.80\pm 0.29\pm 0.53$\ &\ $3.94\pm 0.25\pm 0.42$\\ \hline
$E_e$, $\nu$~reconstruction  (88M $B\bar B$) \cite{b2u_nurec_babar} && \\
$\tilde{E}_e>2.0$~GeV, $\tilde{s}_h^{max}<3.5$~GeV$^2$  & $3.54\pm 0.33\pm 0.34$&$3.95\pm 0.26^{+0.63}_{-0.49}$\\
Full reconstruction tag ($p_\ell^*>1$~GeV/$c$) &&\\
(232M $B\bar B$) \cite{b2u_FR_babar} && \\
$M_X<1.7$~GeV/$c^2$, $q^2>8$~GeV$^2/c^2$ & $8.7\pm 1.3\pm 0.1$&$4.65\pm 0.34\pm 0.49$\\
\hline\hline 
\end{tabular}
\caption{New results on $|V_{ub}|$ from inclusive semileptonic decays.  $\Delta{\cal B}$ is the branching fraction in the restricted kinematic region examined for each result.}
\label{tab:a}
\end{table}

The $|V_{ub}|$ values reported from the analyses discussed above are summarized in Table~\ref{tab:a}.
Even though they are nominally from the same decays, the $|V_{ub}|$ results cannot be compared directly because the experimental partial rates are obtained in different kinematic regions and translated to $|V_{ub}|$ by different schemes and parameter values.  
The HFAG
has been working toward a unified determination of $|V_{ub}|$, based on  all $\Upsilon$(4S) results under a common theoretical framework.
As of this conference, preliminary versions are available at  \{http://www.slac.stanford.edu/xorg/hfag/semi/lp05/lp05.shtml\}.
While the partial rates are all interpreted in the shape function scheme using LNP to obtain  $|V_{ub}|$, there is not yet a consensus on the form and scheme used to obtain the parameters of the shape function.
In the current HFAG summary, heavy quark parameters are obtained in the kinetic scheme by fitting $(i)$ $b\to c\ell\bar\nu$ results only, and $(ii)$ $b\to c\ell\bar\nu$ and $b\to s\gamma$ results.
Figure~\ref{fig:shapefun}(right) shows the results and 1$\sigma$ error contours, which are  straightforwardly translated to the parameters of the shape function scheme to give $m_b$(SF)$=4.60\pm 0.04~{\rm GeV}/c^2$ and $\mu_\pi^2$(SF)$=0.20\pm 0.04~{\rm GeV}^2/c^2$.
From the fit to $(ii)$, HFAG obtains:
\begin{eqnarray*}
|V_{ub}| = (4.39\pm .20 (exp)\pm 0.27 (m_b, th.))\times 10^{-3}.
\end{eqnarray*}
Thus far HFAG has not fitted $B\to X_s\gamma$ data to obtain the shape function directly.  

\subsection{Exclusive semileptonic decays}
Experimentally, the reconstruction of exclusive semileptonic decays such as $B\to\pi\ell\bar\nu$ and $B\to\rho\ell\bar\nu$ is less subject to uncertainty than inclusive reconstruction.
The reporting in the past year of the first unquenched lattice QCD (LQCD) calculations of these decays\cite{pilnu_lqcd1,pilnu_lqcd2} as well as improved calculations using light-cone sum rules (LCSR)\cite{pilnu_lcsr1, pilnu_lcsr2} has brought these exclusive modes into play as contributors to $|V_{ub}|$.
These two methods have complementary regimes of theoretical robustness, with LQCD considered most reliable at high $q^2$ and LCSR being valid at low $q^2$.

\begin{figure}
  \includegraphics[height=5.cm]{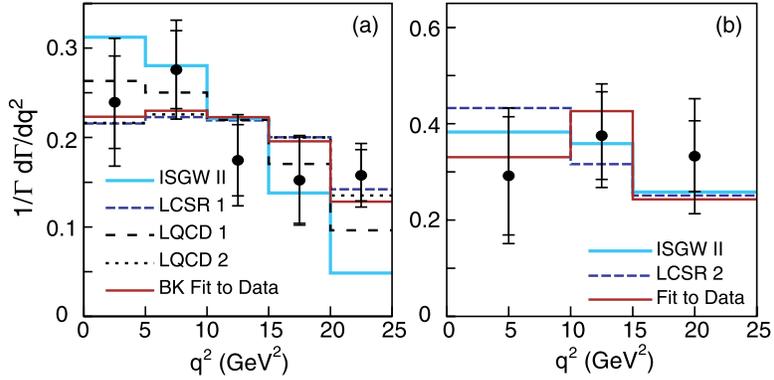}
\caption{Babar \protect\cite{pilnu_nu_babar}: Distributions in $q^2$ of (a)~$B\to\pi\ell\bar\nu$ and (b)~$B\to\rho\ell\bar\nu$ from a neutrino reconstruction analysis.  Data points are shown with error bars; theory references are ISGW2\protect\cite{isgw2}, LCSR1\protect\cite{pilnu_lcsr1}, LQCD1\protect\cite{pilnu_lqcd1}, LQCD2\protect\cite{pilnu_lqcd2}.  }
  \label{fig:pilnu_babar}
\end{figure}

Figure~\ref{fig:pilnu_babar} shows the distributions in $q^2$ of $B\to\pi\ell\bar\nu$ and $B\to\rho\ell\bar\nu$ decays as measured by Babar\protect\cite{pilnu_nu_babar} with neutrino reconstruction.  Superimposed are predictions from various theories.

Another method of reconstructing these decays involves tagging with exclusive semileptonic decays $B\to D^{(*)}\ell\bar\nu$ and requiring kinematic consistency between the two (partially) reconstructed $B$ candidates.
Babar has reported new results for $\bar B^0\to\pi^+\ell^-\bar\nu$\cite{pilnu_D_+_babar} with 232M~$B\bar B$ events and
$\bar B^-\to\pi^0\ell^-\bar\nu$\cite{pilnu_D_0_babar}, with 83M~$B\bar B$ events.
The most recent Belle result,  for $\bar B^0\to\pi^+\ell^-\bar\nu$ and $\bar B^0\to\rho^+\ell^-\bar\nu$ in 152M~$B\bar B$ events, is reported in \cite{pilnu_D_belle}.

The HFAG derives average values based on measurements of $B\to\pi\ell^-\bar\nu$ with $q^2>16$~Gev$^2$  for each of the two LQCD results: $|V_{ub}|$=$(3.75\pm 0.27^{+ 0.64}_{-0.42})\times 10^{-3}$ {\cite{pilnu_lqcd2}} and $|V_{ub}|$=$(4.45\pm 0.32^{+ 0.69}_{-0.47})\times 10^{-3}$ {\cite{pilnu_lqcd1}}, where for each the first error is experimental and the second is due to the normalization uncertainty in the form factor calculation.
These two values mirror the trend that can be seen from the tables shown in the reports of these results -- while there appears to be a spread, the results are consistent within the theoretical uncertainties.  

\section{Summary}
\begin{figure}[b]
  \includegraphics[height=5.08cm]{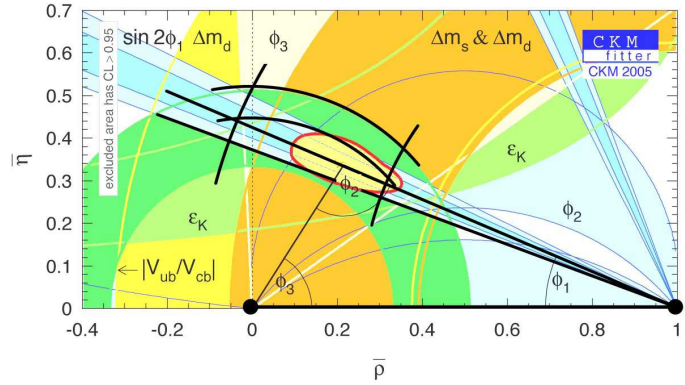}
\caption{Results in the context of the Unitarity Triangle:  Superposition of global fit to $\rho$ and $\eta$ after Winter 2005 conferences \cite{ckmfitter} and $\pm 1\sigma$ zones (black lines) from Summer 2005 measurements of sin2$\phi_1$,  $|V_{td}|$ presented here (arcs centered at (1,0)), and $|V_{ub}|$ presented here  (arcs centered at (0,0)).  As can be seen, the 1$\sigma$ zone from constraints on $|V_{ub}|$ is of comparable size to that from constraints on  sin2$\phi_1$.}
  \label{fig:ckm}
\end{figure}
Since the $B$ factories began operations in 1999,  three quarters of a billion $B$ pair events have been collected by the two experiments, Belle and Babar.  
While the measurements of the CP phases, the angles of the Unitarity Triangle, have received much attention, sufficiently precise measurements of the sides of the triangle will be able to overconstrain its shape and enhance our ability to discern evidence for new physics.  
The past few years have seen significant advances in precision, most notably as applied to the rarer modes that are sensitive to $|V_{td}|$ and $|V_{ub}|$.
In particular, the first observation of $b\to d\gamma$ transitions has already led to a measurement of $|V_{td}|$ with errors of around 13\%.  
The measurement of $|V_{cb}|$ has focused on  moments of the inclusive leptonic energy and hadronic invariant mass spectra, with resulting precisions at the level of 2\%.
Stimulated by large datasets and active communications between experimentalists and theorists, there are now a multitude of measurements of $|V_{ub}|$ involving many  methods and calculations with sufficiently small experimental and theoretical uncertainties ($<10\%$) that possible inconsistencies are being revealed. 
The HFAG is working toward a uniform presentation to allow meaningful comparisons, and this area promises to continue to be an active area that will likely result in continued improvements over the next few years.  
The constraints on $(\rho,\eta)$ from the $|V_{td}|$ and $|V_{ub}|$ results presented here as well as from the most recent measurements of sin2$\phi_1$\cite{abe_lp05} are sketched over the Winter 2005 compilation of the CKM Fitter group\cite{ckmfitter} in
Figure~\ref{fig:ckm}.
It can be seen that these constraints are competitive with those obtained through CP measurements.

\begin{theacknowledgments}
I thank the conference organizers for a most pleasant and stimulating conference.  Thanks are also due to A.~Limosani for consultations on $|V_{ub}|$.
\end{theacknowledgments}


\bibliographystyle{aipproc}   % if natbib is available

\def\etal{{\it et al.}}
\def\PL{{\it Phys. Lett.} }
\def\PLB{{\it Phys. Lett.} {\bf B}}
\def\PRL{{\it Phys. Rev. Letters} }
\def\PRD{{\it Phys. Rev.} {\bf D}}

\end{document}